\documentclass[mathleft]{an}
\usepackage{array,graphicx}
\usepackage{times}
\def \f#1{\ensuremath{\vec{#1}}}
\def \n  {\vec{\nabla}\!}
\def \p  {\partial}
\def \t  {\times}
\def \Re {\mathrm{Re}}
\def \Rm {\mathrm{Rm}}
\def \Ha {\mathrm{Ha}}
\def \Pm {\mathrm{Pm}}
\def \Om {\Omega}
\def \muo{\hat{\mu}_\Omega}
\def \mub{\hat{\mu}_{\mathrm B}}
\def \nuT{\nu_{\rm T}}
\def \nut{\nu_{\rm T}/\nu}

\begin{document}

\Pagespan{1}{}
\Yearpublication{2008}%
\Yearsubmission{2008}%
\Month{0}%
\Volume{000}%
\Issue{0}%

\title{Toroidal field instability and eddy viscosity in Taylor-Couette flows}

\author{M. Gellert\fnmsep\thanks{{Corresponding author: mgellert@aip.de}} \and G. R\"udiger\newline}
\titlerunning{Eddy viscosity in Taylor-Couette flows}
\authorrunning{M. Gellert \& G. R\"udiger}
\institute{Astrophysikalisches Institut Potsdam, An der Sternwarte 16, D-14482 Potsdam, Germany} 

\received{2008}
\accepted{2008}
\publonline{2008}


\abstract{%
         Toroidal magnetic fields subject to the Tayler instability can transport angular momentum. We show that the 
         Maxwell and Reynolds stress of the nonaxisymmetric field pattern depend linearly on the shear in the
         cylindrical gap geometry. Resulting angular momentum transport also scales linear with shear. It is directed outwards
         for astrophysical relevant flows and directed inwards for superrotating flows with ${\rm d}\Om/{\rm d}R>0$. We define an 
         eddy viscosity based on the linear relation between shear and angular momentum transport and show that its maximum for given 
         Prandtl and Hartmann number depends linear on the magnetic Reynolds number $\Rm$. For $Rm\simeq1000$ the eddy viscosity is 
         of the size of $30$ in units of the microscopic value.
         }

\maketitle

\section{Introduction}

Magnetic field configurations in astrophysical objects occur in wide variety. They play an essential role in the formation of flow pattern, are source and
result of dynamo processes and enhance transport processes, where especially transport of angular momentum is of interest here. Crucial for magnetic field 
evolution is the
interplay between toroidal and poloidal fields. In a differentially rotating environment for instance a weak poloidal field is wind-up into a strong toroidal 
one. A strong toroidal field on the other hand can become unstable due to the Tayler instability (\cite{tayler_1957,tayler_1973,vanda72}) or the Azimuthal 
magnetorotational instability (AMRI, see \cite{rued07}). As result also new components of a poloidal field are generated (\cite{gellert07}). To study in more detail
the occurrence of the Tayler instability (TI) and especially the connected transport of angular momentum, we use a model in a cylindrical gap, that starts with 
an already dominating toroidal field and look for the fields instability in a differentially rotating environment.\\
Tayler instability mainly leads to nonaxisymmetric fields. This gives raise to the idea it could be part
of a new dynamo mechanism caused only by the magnetic field (\cite{spruit99}). Even if there is a ongoing discussion if such a dynamo could really 
work (\cite{braith06,zahn07,gellert08,rincon08}), the TI might play an essential role for mixing of chemicals (\cite{rued08}), transport of angular 
momentum and resulting flattening of the
rotation profile of stars. Estimates for angular momentum transport based on the so-called 'Tayler-Spruit' dynamo are in use as input for star evolution 
models (\cite{maeder05,yoon06}). Putting such estimates on more robust basis is the aim of this work. It uses nonlinear 3D simulations of the Tayler instability
in a cylindrical gap to compute the actual angular momentum transport 
\begin{equation}
 T=\langle u_R' u_\phi' - \frac{1}{\mu_0\, \rho} B_R' B_\phi'\rangle,
\label{eq_t}
\end{equation}
where field fluctuations are defined as nonaxisymmetric contributions of the whole velocity and magnetic field in cylindrical coordinates $(R, \phi, z)$. 
Further on it is shown that $T$ scales linear with the differential rotation, that is angular momentum transport can be expressed in terms of a positive 
turbulent or eddy viscosity $\nu_{\rm T}$. We find that for given shear and magnetic field strength always exists a maximum value of the eddy viscosity, 
which grows linearly with the magnetic Reynolds number $\Rm$.

\section{Basic equations and numerics}

We use the magnetohydrodynamic Fourier spectral element code described by Gellert, R\"udiger \& Fournier (2007). With this approach we solve the 3D MHD equations 
\begin{eqnarray}\label{eq_ns}
\lefteqn{ \p_t \f{u}  + (\f{u}\cdot\n)\f{u} = -\n p + \n^{\,2} \f{u} + \frac{\mathrm{Ha}^2}{\mathrm{Pm}} (\mathrm{rot}\f{B}) \t \f{B},} \\
 \lefteqn{ \p_t \f{B}  = \frac{1}{\mathrm{Pm}} \n^{\,2} \f{B} + \mathrm{rot}(\f{u}\t\f{B}),}\\
 \lefteqn{\mathrm{div}\f{u}  = 0, \qquad \mathrm{div}\f{B} =0} 
\end{eqnarray}
for an incompressible medium in a Taylor-Couette system with inner radius $R_\mathrm{in}$ and outer radius $R_\mathrm{out}$. Free parameters are 
the Hartmann number 
\begin{equation}\label{eq_ha}
 \mathrm{Ha} = B^0_\mathrm{in}\sqrt{\frac{R_\mathrm{in} D}{\mu_0\rho\nu\eta}}
\end{equation}
and the magnetic Prandtl number $\mathrm{Pm} = \nu/\eta$. Here $\nu$ is the viscosity of the fluid and $\eta$ its magnetic diffusivity.  
The Reynolds number is defined as $\mathrm{Re}=\Omega_\mathrm{in} R_\mathrm{in} D / \nu$ with  $D=R_\mathrm{out}-R_\mathrm{in}$ (unit of length) and the
angular velocity of the inner cylinder $\Omega_\mathrm{in}$. Unit of velocity is $\nu/D$ and unit of time the viscous time $D^2/\nu$.

The solution is expanded in $M$ Fourier modes in the azimuthal direction. This gives rise to a collection of meridional problems, each of which is 
solved using a Legendre spectral element method (see e.g. \cite{deville_02}).  Either $M=8$ or $M=16$ Fourier modes are used, three elements in 
radial and eighteen elements in axial direction. The polynomial order is varied between $N=10$ and $N=16$. With a semi-implicit approach consisting 
of second-order backward differentiation formula and third order Adams-Bashforth for the nonlinear forcing terms time stepping is done with 
second-order accuracy (\cite{fournier04}).

Boundaries at the inner and outer walls are assumed to be perfect conducting. In axial direction we use periodic boundary conditions to avoid 
perturbing effects from solid end caps. The periodicity in $z$ is set to $\Gamma=6D$.

The simulations start with an initial angular velocity profile
\begin{equation}\label{eq_couette}
 \Omega(R) = a + \frac{b}{R^2},
\end{equation}
the typical Couette profile,
with
\begin{equation}
a = \frac{\hat{\mu}_\Omega - \hat{\eta}^2}{1-\hat{\eta}^2}\Omega_\mathrm{in}, \qquad b = \frac{1-\hat{\mu}_\Omega}{1-\hat{\eta}^2}R^2\Omega_\mathrm{in},
\end{equation}
$\hat{\eta}=R_\mathrm{out}/R_\mathrm{in}=0.5$ and $\hat{\mu}_\Omega = \Omega_\mathrm{out}/\Omega_\mathrm{in}$.
Following the Rayleigh stability criterion $\p_R (R^2\Omega)^2>0$ 
a configuration with $\muo>0.25$ is hydrodynamically stable. This is always fulfilled in our computations. In addition to the shear flow a toroidal 
external field $\f{B_\mathrm{ext}}=(0,B^0,0)$ is applied. The second component 
$B^0$ has a radial profile similar to the flow with a component that is connected to a current only within the inner cylinder wall and a second with
a current also within the fluid domain:
\begin{equation}\label{bext_prof}
 B^0(R) = a_B R + \frac{b_B}{R}.
\end{equation}
Thus the external field than can be characterized by the number
\begin{equation}\label{mub_relation}
 \hat{\mu}_B = \frac{B^0_\mathrm{out}}{B^0_\mathrm{in}} = \frac{a_B R_{\mathrm{out}} + b_B R_{\mathrm{out}}^{-1}}{a_B R_{\mathrm{in}} + b_B R_{\mathrm{in}}^{-1}}.
\end{equation}
The current-free field is given by $\hat{\mu}_B=0.5$. Values below lead to a field connected with constant negative current $j_z$ 
within the flow and larger values to a positive current. The TI occurs in the case that
\begin{equation}\label{eq_ti}
\frac{\rm d}{{\rm d}R} (RB_\phi^2) > 0.
\end{equation}
In the following $\mub$ is fixed to the value $\mub=1$. This means a nearly constant field along radial direction, which is unstable after 
the stability condition (\ref{eq_ti}).

The initial magnetic field in the simulations consists of small 
random perturbations (white noise with an amplitude of $10^{-6}B^0_\mathrm{in}$).

\section{Results}
\subsection{Tayler instability}
\begin{figure}
\includegraphics[width=0.45\textwidth]{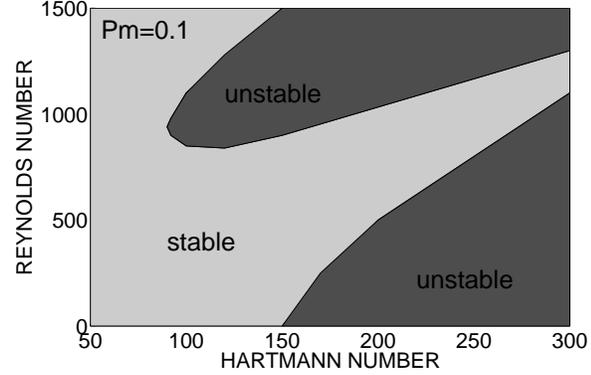}
\caption{Stability map for TI with $\Pm=0.1, \muo=0.5$ and $\mub=1.0$. There exist two separated regions where the instability occurs.}
\label{fig_stab}
\end{figure}

The TI needs basically two conditions to be fulfilled to work. This is on one hand a strong enough magnetic field for given values of $\muo$ and
$\mub$. And, on the other hand, it needs a not to high Reynolds number, because otherwise the instability is suppressed by the shear. In R\"udiger et al. (2007)
stability maps for different Prandtl numbers resulting from a linear analysis are presented. The structure of all these maps for $\Pm\leq1$ is roughly the same. 
For the case of $\Pm=0.1$ such a stability map is shown in Fig. \ref{fig_stab} with values resulting from computations with the full 3D code. It differs only 
slightly from the linear results. Remarkable is the separation of the unstable region by a stable island. Nevertheless the appearance of the TI is in both 
regions the same.
The most unstable mode is the nonaxisymmetric mode $m=1$. And in both regions exist a upper boundary for the Reynolds number where the instability disappears. With
smaller $\Pm$ the critical Reynolds numbers grow. For $\Pm=1$ it is roughly $\Re=450$ for Hartmann number $\Ha=150$, for $\Pm=0.1$ already $\Re=1500$.

\subsection{Definition of eddy viscosity}\label{sec_def_nut}
For the calculation of the Maxwell and Reynolds stress we define mean fields as quantities averaged in azimuthal direction:
\begin{equation}
 \langle \f{u}\rangle = \frac{1}{2\pi} \oint \f{u}\, {\rm d}\phi, \quad \langle \f{B}\rangle = \frac{1}{2\pi} \oint \f{B}\, {\rm d}\phi.
\end{equation}
This regards nonaxisymmetric parts as fluctuations $\f{u'}=\f{u}-\langle \f{u}\rangle$ and $\f{B'}=\f{B}-\langle \f{B}\rangle$ respectively. The normalized 
equation for angular momentum transport (eq.\ref{eq_t}) then reads
\begin{equation}
 T\simeq \langle u_R' u_\phi'\rangle - \frac{\Ha^2}{\Pm} \langle B_R' B_\phi'\rangle.
\label{eq_tn}
\end{equation}
The question arising now is whether angular momentum is transported outwards, e.g. $T>0$ for ${\rm d}\Om/{\rm d}R<0$. If this is true, an eddy viscosity $\nu_{\rm T}$
can be introduced by defining
\begin{equation}
 T= - \nu_{\rm T} R \frac{{\rm d}\Om}{{\rm d}R}
\label{eq_nut}
\end{equation}
with positive $\nu_{\rm T}$.

\begin{figure}
\includegraphics[width=0.45\textwidth]{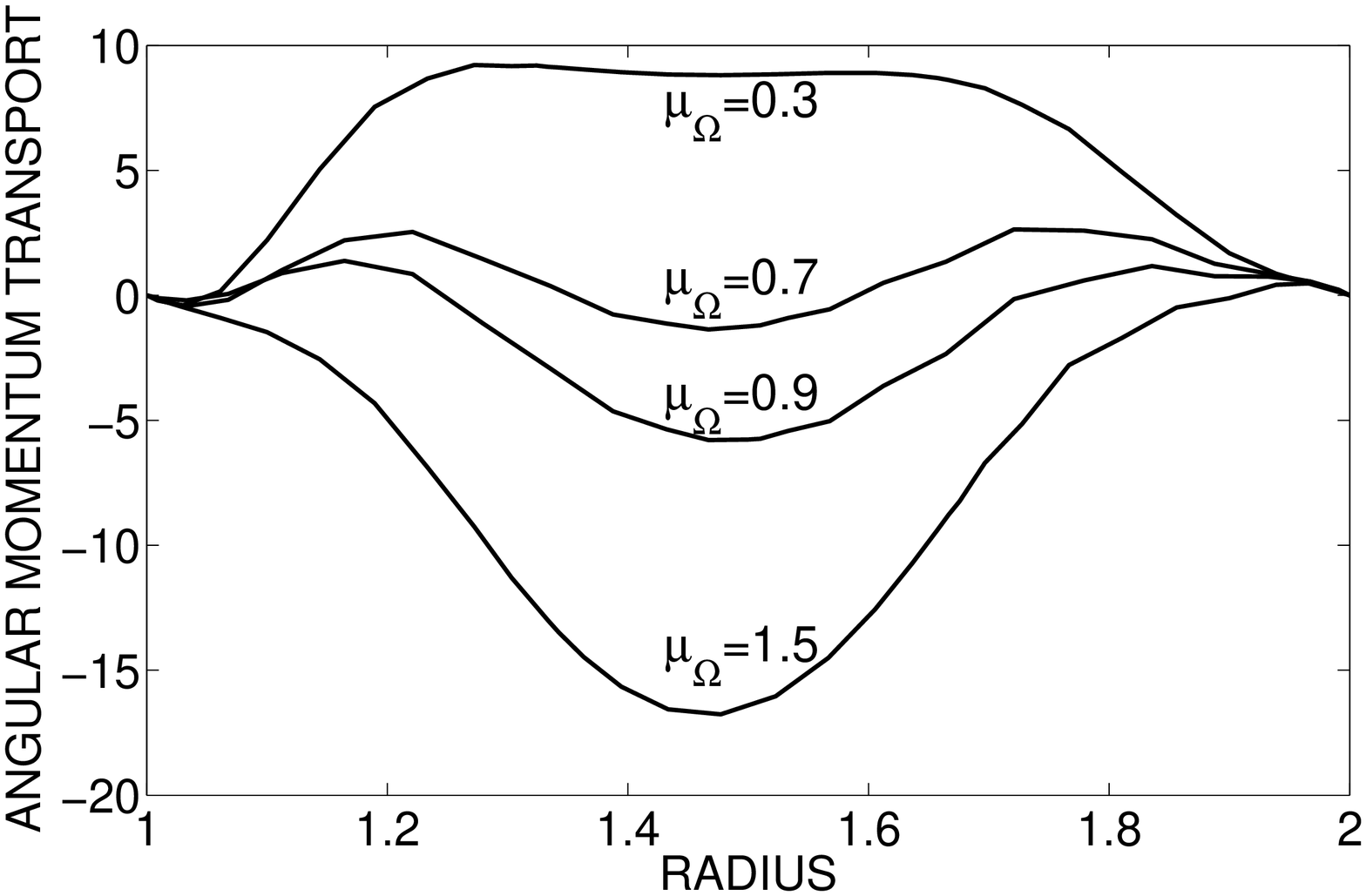}
\caption{Radial profile of angular momentum transport $T$ for several differential rotation profiles $\muo$. Reynolds number is $\Re=30$
         and Hartmann number $\Ha=190$. $T$ is mainly positive for subrotation $d\Om/{\rm d}R<0$ and becomes negative for superrotation $d\Om/{\rm d}R>0$.}
\label{fig_mu_prof}
\end{figure}

Radial profiles of $T$ for $\Pm=1$, $\Re=30$, $\Ha=190$ and $\mub=1$ found in our nonlinear simulations are shown in Fig. \ref{fig_mu_prof}. We find indeed that the 
value of $T$ depends on $\muo$. The profile is everywhere positive for $\muo<0.65$ and everywhere negative for $\muo>1.1$. For values
in between there are regions where $T$ is positive or negative. Taking a radially averaged value of $T$ shows, that this average is positive for
$\muo<0.75$ and negative for larger $\muo$.

\begin{figure}
\includegraphics[width=0.45\textwidth]{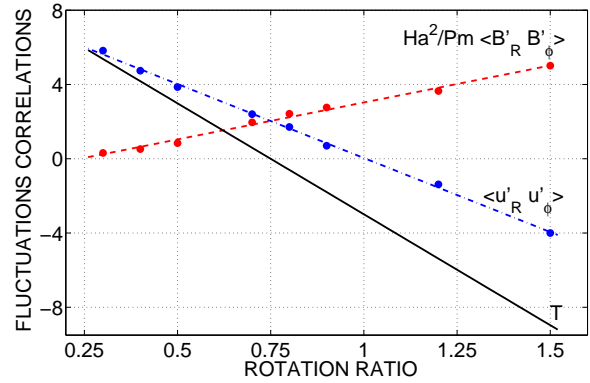}
\caption{Scaling of angular momentum transport, Maxwell and Reynolds stress with the shear, expressed as rotation ratio $\muo$ of inner and outer cylinder. Positive
$T$ means transport of angular momentum outwards.}
\label{fig_muom}
\end{figure}

It further on reveals a linear dependence between $T$ and $\muo$ in the whole range of $0.3<\muo<1.5$ (see Fig. \ref{fig_muom}). The kinetic part decreases
with increasing $\muo$, it is zero for solid rotation as one would expect. Contrary the magnetic part increases with $\muo$ and is positive. The linear fit
crosses zero at the Rayleigh line. There the pure hydrodynamic instability dominates and makes numerical checks impossible. The effect, that $T$ is not
zero for solid rotation but has a contribution $T^{\ast}\sim\Om$ is called $\Lambda$-effect (\cite{rued89,kit93}). Its appearance her is not yet investigated 
in detail.\\
Because of the linear scaling of $T$ with $\muo$, the introduction of an eddy viscosity as defined by eq. (\ref{eq_nut}) is legitimate. If we 
substitute $R {\rm d}\Om / {\rm d}R$ by using $\muo$ as
\begin{equation}
\frac{{\rm d}\Om}{{\rm d}R} = (1-\muo)\frac{R \Om_\mathrm{in}}{D}
\end{equation}
in eq. (\ref{eq_nut}) and using the slope of the linear function \mbox{$T=T(\muo)$} plotted in Fig. \ref{fig_muom}, we get an eddy viscosity in units of the 
microscopic viscosity of $\nut = 0.39$ for the example with $\Ha=190$ and $\Re=30$. This is rather low and due to the fact that $\Re$ is chosen to be 
very small. In the astrophysically more relevant regime, where $\Re$ is (much) larger than $\Ha$ the eddy viscosity increases rapidly too.\\ 
Before describing the scaling of $\nuT$ with increasing Reynolds number and decreasing Prandtl number a remark regarding the size of the 
Maxwell and Reynolds stress is appropriate. As shown in Fig. \ref{fig_muom} both values are of the same order. With increasing $\Re$ and $\Ha$ the magnetic part 
becomes more and more dominating. So the turbulent
viscosity is mainly due to the magnetic field action. On the other hand, the velocity correlation stays on this low level. Arguing that the mixing
of chemicals in stars is connected only to the kinetic part of the correlation tensor, mixing due to the TI would be rather weak (\cite{rued08}) and 
in agreement with observations of lithium concentration in solar type stars (\cite{brott}).

\subsection{Reynolds and Hartmann number dependence of eddy viscosity}
\begin{figure}
\includegraphics[width=0.45\textwidth]{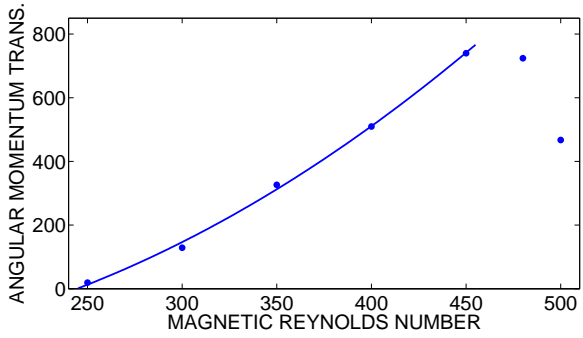}
\caption{Angular momentum transport for $\Pm=1$, $\Ha=190$ with $\mub=1$ and $\muo=0.5$. The rotational stabilization for large $\Re$ leads to the 
         existence of a maximum value. The fitted line, a parabola, suggests a relation $T\sim {\rm Rm}^2$.}
\label{fig_angu}
\end{figure}

To study the influence of turbulent viscosity for instance in star evolution models, one would like to have a relation at hand connecting the size of $\nut$ with 
given (magnetic) Reynolds and Hartmann numbers. This are usually large values and are not reachable with our 3D simulations, thus a direct computation
is impossible at the moment. Nevertheless, in a first step we want to give a scaling relation of $\nuT$ depending on $\Re$ and/or $\Ha$ for $0.1\leq\Pm\leq 1$.\\
As already mentioned the TI for given Hartmann number occurs only in a certain range for $\Re$. Along a line of constant $\Ha$ starting from the lower critical
value of $\Re$ the instability first becomes stronger in a sense, that the angular momentum transport increases. It reaches a maximum and 
fades away towards the upper critical $\Re$. This behavior is shown in Fig. \ref{fig_angu} for $\Pm=1$ and $\Ha=190$. The fit with a second order polynomial 
suggests a $T \sim {\rm Rm}^2$ scaling for the increasing part.

\begin{figure}
\includegraphics[width=0.45\textwidth]{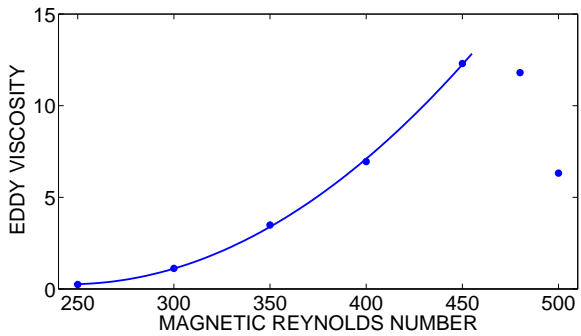}
\caption{Same as in Fig. \ref{fig_angu} but for the normalized eddy viscosity $\nut$.}
\label{fig_nut}
\end{figure}

The calculation of $\nuT$ from $T$ using eq. (\ref{eq_nut}) gives the values plotted in Fig. \ref{fig_nut}. The dependence seems to be the same as for T:
$\nuT \sim {\rm Rm}^2$ if far enough from the upper stability boundary. The maximum value in this example ($\Ha=190$) for the eddy viscosity is 
${\rm max}(\nut) = 12.3$. It grows further when $\Ha$ is increased and
reaches an absolute maximum of the order of 30 for $\Ha\approx 500$ for $\Pm=1$. For Prandtl numbers below $\Pm=1$, this tendency to reach a saturation 
value is not yet visible in the with our code achievable range for the Reynolds and Hartmann number respectively. The several computed values are listed 
in tab. \ref{tab1}. Because the Reynolds number grid is rather coarse, the given maxima might be slight underestimations of the real maxima, 
which reside between the given value and the first lower point towards the upper stability boundary.

\begin{table}
\caption{\label{tab1} Eddy viscosity $\nuT$ and the belonging Reynolds numbers for which $\nut$ reaches its maximum
         for given Prandtl number and Hartmann number.}
\setlength{\extrarowheight}{4pt}
\begin{tabular}{m{1.2cm}m{1.2cm}m{1.5cm}m{1.2cm}}
\hline
$\Pm$ & $\Ha$ &  max$(\nut)$ &  $\Re$ \\
\hline\hline
0.1 & 150 &  3.8  & 1300 \\
\hline
0.4 & 100 &  5.7  &  550 \\
0.4 & 150 &  9.2  &  850 \\
0.4 & 250 & 17.6  & 1700 \\
\hline
1.0 & 100 &  5.2  &  320 \\
1.0 & 190 & 12.3  &  460 \\
1.0 & 400 & 29.1  & 1100 \\
1.0 & 500 & 30.4  & 1200 \\
\hline
\end{tabular}
\end{table}

The next question we attack is if it is possible to find a relation between the eddy viscosity maxima and the parameters they depend on, and how such a relation 
will look like. The first idea is that such a relation should reflect the influence from all free parameters, thus a quantity depending on magnetic field, 
shear and Prandtl number. Thereby on aspect simplifies the problem: we deal only with the maxima. For them it is equivalent to use $\Re$ 
or $\Ha$, both are not independent. So the relation in question might hide the dependence on the third parameter. Two well-known parameters, the Lundquist 
number $S$ and the magnetic Reynolds number $\Rm$ combine the influence from two physical properties of the system. Both are first candidates to check. 
The plot of $\nuT(S)$ does not reveal a direct relation between eddy viscosity and Lundquist number and is not shown here.

\begin{figure}
\includegraphics[width=0.45\textwidth]{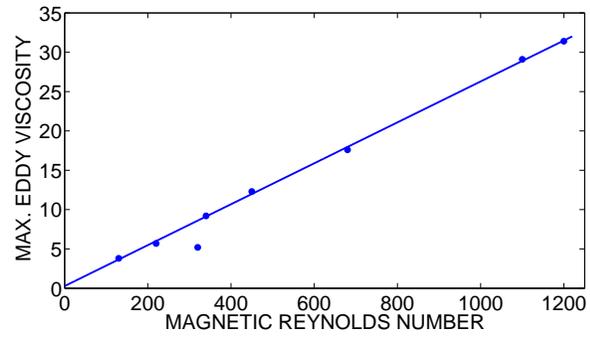}
\caption{Linear dependence of eddy viscosity maxima on the magnetic Reynolds number for the investigated Prandtl number range $0.1\leq\Pm\leq1$.}
\label{fig_rm}
\end{figure}

On the other hand a plot of $\nuT$ as function of $\Rm$ shows indications of the relation in question.
Using the values of $\nut$ listed in tab. \ref{tab1} gives the data points plotted in Fig. \ref{fig_rm}. Those points reveal a linear dependence
\begin{equation} \label{eq_rm}
 \nut = 0.026\, \Rm + 0.45 
\end{equation}
between eddy viscosity maxima and magnetic Reynolds number independent on Prandtl number. The one exceptional point much below the line for $\Rm=320$ is 
due to the fact, that this point, belonging to $\Ha=100$, is to close to the left stability boundary (see fig. \ref{fig_stab}, for $\Pm=1$ the Hartmann 
number range is very the same).  Lets remind the reader that $\nuT$ here means the maxima of the eddy viscosity produced by the TI for given Prandtl and 
Hartmann (or equivalent Reynolds) number. As visible in Fig. \ref{fig_nut} the linear relation $\nuT(\Rm)$ is not true in general for arbitrary combinations 
of Hartmann and Reynolds numbers.

The derived relation (\ref{eq_rm}) agrees also quite well with the resulting value for
the eddy viscosity described in subsection \ref{sec_def_nut} using the linear dependence between $T$ and the differential rotation parameter $\muo$. Indeed,
going to higher Hartmann numbers leads to an increase of the eddy viscosity to a value of $\nut=1.1$, which is nearly exactly the value relation (\ref{eq_rm}) 
predicts for $\Rm=30$.

The scaling with $\Rm$ reflects the raising influence of the magnetic field fluctuations for faster rotation. It is derived for nearly constant magnetic 
fields ($\mub=1$) and a Prandtl number between $\Pm=0.1$ and $\Pm=1$. If we assume it holds 
also for smaller Prandtl numbers, one could expect for instance in young, fast rotating stars with $\Rm\approx1000$ a significant transport of angular
momentum and eddy viscosities of the order of $30$. For a sun-like star with $\Rm\approx150$ it would be a weak effect with $\nut=5$.

\section{Conclusions}
Strong toroidal fields under the influence of a differentially rotating medium can become unstable to nonaxisymmetric perturbations. This instability, the Tayler
instability, appears only if the shear is not to strong. In cylindrical gap geometry for Prandtl numbers $0.1\leq\Pm\leq1$ there is always an upper boundary
for the Reynolds number above which the instability is suppressed. The smoothing action of differential rotation dominates. Additionally it needs the crossing 
of a threshold for the strength of the magnetic field, for a nearly constant toroidal field with $\mub=1$ this is of the order of $\Ha\approx100$. If our
astrophysical object in mind provides such requirements, the TI leads to nonaxisymmetric magnetic field pattern.

Using the nonaxisymmetric parts as fluctuations, the resulting angular momentum transport scales linear with the differential rotation, it is positive for
astrophysical relevant configurations with $\muo\leq0.75$, e.g. angular momentum is transported outwards. This gives raise to the introduction of an eddy
viscosity which causes the additional turbulence in the underlying medium.

For a fixed magnetic field there is always an upper and lower critical Reynolds number in between the instability can be observed. And there is also a maximum
value of angular momentum transport and eddy viscosity. Taking the maxima of the latter we find its values follow a linear dependence on the magnetic Reynolds 
number. A direct influence of the field strength is hidden behind the fact, that with the Reynolds number also the Hartmann number is fixed
for the point where the eddy viscosity has its maxima. 

The linear relation $\nuT(\Rm)$ is derived for a simplified model with fixed toroidal field and without any influence from density stratifications for 
rather high Prandtl numbers $0.1\leq\Pm\leq1$. Although, imagine it hold also for Prandtl numbers one or two orders of magnitude less, this relation could be
used to estimate eddy viscosities usable as input for star evolution models. For instance for a young, fast rotating sun it gives values around $\nut=30$.


\end{document}